# High-Q trenched aluminum coplanar resonators with an ultrasonic edge microcutting for superconducting quantum devices

E.V. Zikiy,[1,2] A.I. Ivanov,[1,2] N.S. Smirnov,[1,2] D.O. Moskalev,[1,2]
V.I. Polozov,[1] A.R. Matanin,[1,2] E.I. Malevannaya,[1] V.V. Echeistov,[1]
T.G. Konstantinova[1] and I.A. Rodionov[1,2,*]

[1]*FMN Laboratory, Bauman Moscow State Technical University, Moscow 105005, Russia*
[2]*Dukhov Automatics Research Institute (VNIIA), Moscow 127055, Russia*
*email: irodionov@bmstu.ru

**ABSTRACT**

Dielectric losses are one of the key factors limiting the coherence of superconducting qubits. The impact of materials and fabrication steps on dielectric losses can be evaluated using coplanar waveguide (CPW) microwave resonators. Here, we report on superconducting CPW microwave resonators with internal quality factors systematically exceeding $5 \times 10^6$ at high powers and $2 \times 10^6$ (with the best value of $4.4 \times 10^6$) at low power. Such performance is demonstrated for 100-nm-thick aluminum resonators with 7-10.5 um center trace on high-resistivity silicon substrates commonly used in quantum Josephson junction circuits. We investigate internal quality factors of the resonators with both dry and wet aluminum etching, as well as deep and isotropic reactive ion etching of silicon substrate. Josephson junction compatible CPW resonators fabrication process with both airbridges and silicon substrate etching is proposed. Finally, we demonstrate the effect of airbridges' positions and extra process steps on the overall dielectric losses. The best quality fa ctors are obtained for the wet etched aluminum resonators and isotropically removed substrate with the proposed ultrasonic metal edge microcutting.

Superconducting CPW microwave resonators are the basic elements of superconducting circuits: quantum processors,[1] quantum-limited parametric amplifiers,[2] quantum memory,[3] photon detectors,[4] and artificial atoms.[33] There are many applications where resonators operating in a single-photon regime are characterized by a significant internal quality factor ($Q_i$) decrease due to dielectric losses in bulk dielectrics and thin interfaces containing two-level systems (TLS)[5,6]. Dielectric losses directly affect the performance of superconducting devices, for example, the relaxation times of qubits.[5,7] CPW resonators internal quality factor at low microwave power ($Qi_{LP}$) depends dominantly on dielectric losses in interfaces: metal-substrate (MS), metal-vacuum (MA) and substrate-vacuum (SA) interfaces.[8,28] It is well known, that the MS interface is dominant[28] and it is generally determined by the choice of metal deposition and substrate cleaning procedures[31]. High $Qi_{LP}$ values approaching $2.0 \times 10^6$ were obtained for TiN[8] and NbTiN[10] CPW resonators. However, thick metal films up to 300 nm and 750 nm respectively were used, which are not applicable for qubits fabrication. The best $Qi_{LP}$ reaching $2.0\text{-}3.0 \times 10^6$ in case of 100 nm thick aluminum film were demonstrated[13] for large footprint CPW resonators (center trace of 24 μm). A silicon substrate etching with Al resonators was implemented in Ref. 29, but with 250 nm thick aluminum the best $Qi_{LP}$ up to $1.8 \times 10^6$ was achieved. Internal quality factor of CPW resonators itself can be increased using new materials compatible with aggressive treatment, thicker superconducting films and larger footprint of resonators leading to lower field intensity. However, it is very hard to integrate them into superconducting qubit circuits fabrication processes. Aluminum technology is still one of the leading platforms for superconducting qubits,[11,12] which requires base sub-150 nm thick Al layer[12,14,16] with optimized footprint resonators (center trace up to 10 μm[17]). Improving aluminum CPW resonators quality requires further technology investigation: ultra-high vacuum Al deposition,[13] advanced substrate cleaning[14], substrate etching,[8,10] and etc.

In this paper, we report on high $Qi_{LP}$ aluminum 100 nm thick compact resonators on etched silicon substrates compatible with superconducting qubits fabrication. We investigate Al metal and Si substrate etching, as well as post treatment steps, in order to reduce the loss on the MA and SA interfaces. Using the proposed technology, we demonstrate internal quality factors at low $Qi_{LP}$ and high $Qi_{HP}$ power exceeding $2.0 \times 10^6$ and $5.0 \times 10^6$ respectively for identical resonators at frequencies ranging from 4.0 to 5.0 GHz. It is fabricated using isotropic substrate etching of optimized footprint compact resonators (10.5 μm center trace and 3.5 μm gap) with both airbridges and without them. The best internal quality factors obtained for the 2.91 GHz resonator are $Qi_{LP} = 4.4 \times 10^6$ and $Qi_{HP} = 1.9 \times 10^7$. We achieve it by introducing isotropic silicon substrate etching with subsequent ultrasonic resonators edge microcutting after aluminum wet etching.

**TABLE I.** CPW resonators comparison; w is the resonator center trace width, gap is the gap between resonator center trace and the ground, $f_0$ is the resonant frequency, Qc is the coupling quality factors between feedline and resonators, and $Qi_{LP}$ are the internal quality factors at low power.

| $f_0$, GHz | w, um | Gap, um | Film | Thickness, nm | Substrate | Substrate etching | Qc, x10^5 | $Qi_{LP}$, x10^5 | Ref. |
|---|---|---|---|---|---|---|---|---|---|
| 5.50 - 6.0 | 24.0 | 24.0 | Al | 100 | Si 100 | - | - | 20.0 -30.0 | 13 |
| 2.75 - 6.41 | 12.0 | 5.0 | NbTiN | 160, 300 | Si 100 | + | 7.0 – 10.0 | 10.0 – 20.0 | 10 |
| 5.0 - 6.0 | 28.0 | 14.0 | TiN | 450, 750 | Si 100 | + | - | 20.0 | 8 |
| - | - | - | Al | 250 | Si 100 | + | 1.5 – 50.0 | 18.0 | 29 |
| 4.50 | - | - | Al | 150 | Si 100 | - | 3.4 | 8.0 | 14 |
| 5.20 - 5.60 | - | - | Nb | - | Si | + | - | 8.4 - 11.8 | 30 |
| 2.91 - 5.0 | 7.0/10.5 | 4.0/3.5 | Al | 100 | Si 100 | + | 2.0 - 4.0 | 16.5 - 44.0 | This work |

After resonators patterning, we fabricate airbridges to suppress parasitic slotline modes[18]. In order to evaluated airbridges influence on $Qi_{LP}$, we measured identical resonators without airbridges, with airbridges over feedline only, and over both resonators and feedline. Using the proposed technology, we are able to reach the highest internal quality factor at low power for compact aluminum CPW resonators[8,10,13,14,29,30] compatibles with superconducting qubit circuits fabrication process (Table 1).

To evaluate the effects of the Al film and Si substrate etching, airbridges fabrication, additional ultrasonic microcutting on $Qi_{LP}$ of the resonators, we fabricated quarter-wave resonators according to the frequency multiplexing scheme[19] on 25x25 mm silicon substrates with further cutting to 5x10 mm chips. There are 12 resonators on each chip with frequencies ranging from 4.0 to 7.0 GHz for devices without substrate etching and 6 resonators with frequencies ranging from 2.5 to 5.0 GHz for devices with substrate etching. All the resonators were designed to have 50 Ohm impedance (center trace widths/gap): 7.0/4.0 μm for resonators without substrate etching and 10.5/3.5 μm for resonators with substrate etching. The widths of the etched resonators are corrected to take into account the change in the effective dielectric permittivity ($\varepsilon_{eff}$[20]) during substrate etching. The coupling quality factor Qc was designed to be $3.0 \times 10^5$, but the experimental values are in the range of $2 \times 10^5$ to $4 \times 10^5$ due to simulation and design issues. A script[16] based on a conformal mapping method was used to evaluate Qc and impedance of the resonators. In order to eliminate frequency dependence, we selected and compared the internal quality factors of the resonators with frequencies ranging from 4.0 to 5.0 GHz only.

For airbridges influence evaluation we used two designs: the first one with 9 airbridges over the feedline only; the second one with both 9 airbridges over the feedline and 4 airbridges evenly spaced over each resonator, which should be enough to eliminate the slotline modes.[18] Optical images of the chips can be found in the supplementary materials.

Figure 1(a) shows the fabrication sequence scheme of resonator chips. We used high-resistivity Si(100) substrates (>10kOhm-cm) for all the samples. Al films were deposited by ultrahigh-vacuum electron-beam deposition system under a base pressure lower than $10^{-10}$ Torr. Before deposition substrates were cleaned in RCA1 solution, followed by HF treatment to remove native oxide and terminate the Si surface with hydrogen. Then we installed Si substrates in the load lock as quickly as possible after cleaning, typically within 10 min. Al films with a thickness of 100 nm were deposited according to the regime used in Ref. 21 and Ref. 32 to form the base metal layer. After photoresist mask spincoating and patterning, Al films were etched either by wet etching in an industrial Aluminum Etching Type A solution (Fig. 1(b) or by dry etching in a $BCl_3/Cl_2$ gas mixture (Fig. 1(c). Then we dry etched the silicon substrate either by Bosch DRIE process[22] with 90 cycles (Fig. 1(d)) or by isotropic RIE process in $SF_6$ gas mixture. During Al etching process the edges of resonators center trace are usually damaged (Fig. 1(b), (c)) by thermal or chemical influence. To remove these damaged metal edges, we optimize our substrate etching processes to get the desired undercut, then by using strong ultrasonic microcutting in isopropyl alcohol, we cut them to obtain high-quality metal edges (Fig. 1(e)). At the final stage, airbridges were formed for a group of resonators according to the technology used in superconducting qubit circuits fabrication.[17]

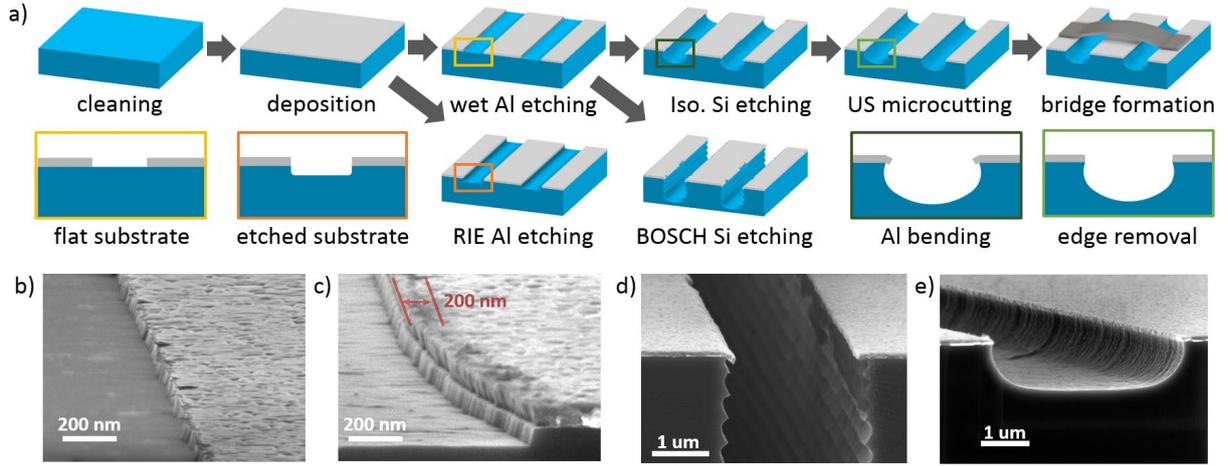

**FIG 1.** (a) Fabrication sequence of resonator chips with different Al film and Si substrate treatment. SEM images of the resonators center trace edges: (b) Al wet etching; (c) Al dry etching; (d) Si substrate Bosch DRIE; the sagging edges of the thin film can be observed; (e) Si substrate isotropic etching followed by ultrasonic microcutting.

After dicing, the chips were mounted in copper sample holders made in according to the recommendations given in Ref. 23, and mounted in a 10 mK stage in the dilution fridge. We used infrared and magnetic shielding to protect our samples against quasiparticles generations[24] and magnetic vortices. We measured the transmission coefficient $S_{21}$ of the resonators with a vector network analyzer (VNA) according to the method described in Ref. 25. A total attenuation of 90 dB was installed on cryostat stages, all the measurements are performed under the temperatures below 50 mK. The input and output lines were equipped with powder infrared filters-eccosorb, as well as low-pass filters. At the output line at 4 K stage, there is an amplifier on a high electron mobility transistor (HEMT). The wiring diagram of a measurement setup for the samples can be found in supplementary materials. We varied the drive power so that the photon population $\langle n_p \rangle$ in the resonator ranged from the single-photon levels up to $10^7$ photons. We experimentally observed, that at the lowest power $Q_{iLP}$ can fluctuate by more than 34% over several hours period due to fluctuations in TLS populations.[9] Here, we present the time-averaged $Q_{iLP}$ values instead of maximum values.

Figure 2(a) shows $Q_{iLP}$ measurements for the CPW resonators grouped by different Al film and Si substrate etching technology. Groups 1a and 1b with the average $Q_{iLP}$ of $6.0 \times 10^5$ and $1.18 \times 10^6$ include resonators obtained by RIE and wet Al etching, respectively, without Si substrate etching. Group 2a with the average $Q_{iLP}$ of $6.1 \times 10^5$ includes resonators obtained by wet Al etching with Si substrate Bosch DRIE. Groups 2b and 2c with the average $Q_{iLP}$ of $1.21 \times 10^6$ and $2.05 \times 10^6$ contain resonators obtained by wet Al etching with Si substrate isotropic etching without ultrasonic edge microcutting and with it, respectively. Figures 2(b), (c), (d) show SEM images of the structure specifics for groups 2a, 2b, 2c. The measurement results of all our samples are shown in the supplementary materials.

One can notice the systematic dependence of $Q_{iLP}$ on the metal and substrate etching processes. We found that the $Q_{iLP}$ of resonators fabricated by wet etching is twice higher compared to our dry etching. We attribute this dependence to the metal-vacuum (MA) and substrate-vacuum (SA) interfaces having significantly lower loss tangents after wet etching than after dry etching. It could be definitely observed, that the surface of resonator center trace is damaged[27] at a distance of about 200 nm from the edge (Fig. 1(c)), which is the area with the highest field intensity. At the same time, it was demonstrated by simulation[28] that the substrate etching by only 10 nm reduces the participation ratio of the metal-air-substrate corners by 50%, while preserving the other participation ratios, which should have a positive effect on the $Q_{iLP}$ level. In our case, we have dry etched the substrate to 80 nm depth, but the $Q_{iLP}$ level is still much lower than in the case of wet etching, where no etching of the substrate took place. We suppose that the reason is a very high concentration of TLS in the damaged region together with the high field intensity.

Bosch DRIE substrate etching allowed the fabrication of resonators with low $Q_{iLP}$ values. The most possible reason is a high TLS concentration in the MA and SA interfaces as a result of incomplete removal of specific Bosch process polymer residues, which could be further cleaned. Isotropic etching of the Si substrate allowed a slight increase in $Q_{iLP}$ compared to the level of wet-etched Al resonators (from $1.18 \times 10^6$ to $1.21 \times 10^6$), but the standard deviation in the group increased significantly. The possible reason is a non-reproducibility of metal edge geometry, which turns out to be "suspended" after etching, which negatively affects MA interface, resonator impedance and resonant frequency. We confirm this assumption by introducing an additional treatment in isopropyl alcohol with ultrasound: the "suspended" metal edge is broken off and the geometry of the resonators is reproduced exactly. With the width of the removed metal being of 600 nm, which is 3 times the width of the Al section damaged during etching, resulting in an almost twofold increase in the average $Q_{iLP}$ to $2 \times 10^6$ while the standard deviation value decreases.

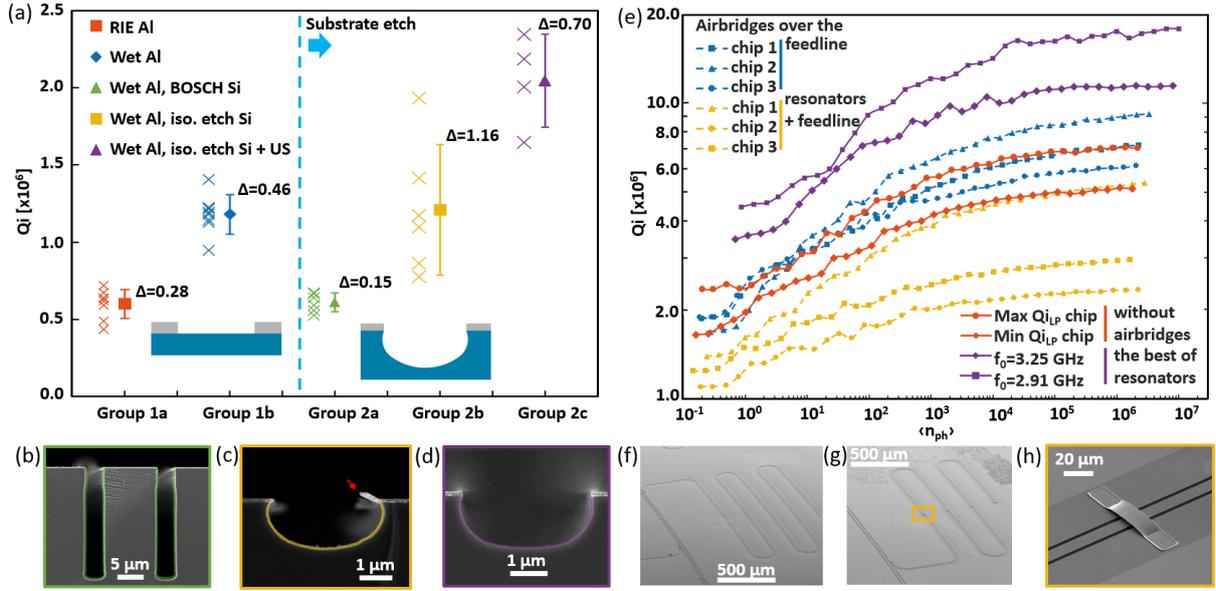

**FIG 2.** (a) Internal quality factor in single-photon regime of resonators grouped by fabrication technological features into groups. Group 1a - RIE Al, without substrate etching; group 1b - wet etching Al, without substrate etching; group 2a - wet etching Al, DRIE Bosch substrate; group 2b - wet etching Al, isotropic substrate etching; group 2c - wet etching Al, isotropic substrate etching, additional ultrasonic microcutting (crosses indicate the average value of $Q_{iLP}$ for each resonator, whereas the error bars indicate the standard deviations and mean value). SEM images of the cross section of the resonators: (b) group 2a, (c) group 2b, (d) group 2c. (e) Dependence of internal quality factor in single-photon regime of resonators with wet etching Al and isotropic etching of substrate with airbridges over feedline (blue lines), with airbridges over feedline and resonators (yellow lines), without airbridges (red lines) in 4 - 5 GHz range and outside this range (violet line) on average photon population in resonator. Lines were added for better visibility. (f) SEM image of the feedline section with bridges; (g) SEM image of the feedline and resonator section with bridges; (h) SEM image of a single bridge on resonator with etching of the substrate in the gap.

Figure 2(e) shows the $Q_{iLP}$ dependences of resonators with wet Al etching and isotropic Si substrate etching with airbridges over the feedline (Fig. 2(f), (h), blue lines), with airbridges over the feedline and resonators (Fig. 2(g), yellow lines), and without airbridges (red lines) on the average photon population in the resonator. The average photons number was determined based on applied power, Qc and the loaded quality factor Ql according to the recommendations from Ref. 26. For the group of resonators without airbridges, only the results with the highest and lowest $Q_{iLP}$ are shown. Figure 2(e) also shows Qi of our best resonators without air bridges at the frequencies 2.91 GHz and 3.25 GHz and $Q_{iLP}$ equal to $4.4 \times 10^6$ and $3.4 \times 10^6$, respectively (violet line). One can notice, that airbridges location directly affects resonator $Q_{iLP}$ (Fig. 2(e)), which is in a good agreement with Ref. 18. The airbridges placed over the feedline does not affect the resonators internal quality factor (it is within the variation of $Q_{iLP}$ for resonators without airbridges).

In summary, we have measured the internal quality factors of 100 nm thick aluminum compact CPW resonators which are compatible with superconducting qubits fabrication route for various base metal and silicon substrate etching processes, as well as post treatment technological step. Wet Al film etching with isotropic Si substrate dry etching followed by the proposed ultrasonic resonators edge microcutting leads to the average $Q_{iLP}$ above $2.0 \times 10^6$, achieved resonators with w = 10.5 μm and $f_0$ = 4.0 – 5.0 GHz. The highest achieved $Q_{iLP}$ value is $4.4 \times 10^6$ for the resonator with w = 10.5 μm and $f_0$ = 2.91 GHz. Finaly, we fabricate high quality factor superconducting CPW resonators with Si substrate etching and airbridges showing that the additional fabrication steps do not result in overall circuit performance decrease. The samples are fabricated at the BMSTU Nanofabrication Facility (Functional Micro/Nanosystems, FMNS REC, ID 74300).

## AUTHOR DECLARATIONS

### Conflict of Interest

The authors have no conflicts to disclose.

### Author Contributions

**Evgeniy Zikiy:** Conceptualization (equal); Formal analysis (lead); Methodology (lead); Investigation (equal); Writing – original draft (lead); Visualization (lead).

**Anton Ivanov:** Methodology (equal); Investigation (equal); Writing – review and editing (supporting).

**Nikita Smirnov:** Investigation (equal); Writing – review and editing (supporting).

**Dmitry Moskalev:** Investigation (supporting); Writing – review and editing (supporting).

**Victor Polozov:** Investigation (equal); Writing – review and editing (supporting).

**Alexey Matanin:** Formal analysis (supporting); Investigation (equal); Writing – review and editing (supporting).

**Elizaveta Malevannaya:** Formal analysis (supporting); Investigation (supporting).

**Vladimir Echeistov:** Formal analysis (supporting); Investigation (supporting).

**Tatiana Konstantinova:** Formal analysis (supporting); Investigation (supporting).

**Ilya Rodionov:** Conceptualization (lead); Investigation (equal); Formal analysis (equal); Writing – original draft (equal); Writing – review and editing (lead); Supervision (lead).

## DATA AVAILABILITY

The data that support the findings of this study are available within the article and its supplementary material.

## REFERENCES

[1] A. Blais, R. S. Huang, A. Wallraff, S. M. Girvin, and R. J. Schoelkopf, "Cavity quantum electrodynamics for superconducting electrical circuits: An architecture for quantum computation." Phys. Rev. A **69**, 062320 (2004).

[2] M. A. Castellanos-Beltran, and K. W. Lehnert, "Widely tunable parametric amplifier based on a superconducting quantum interference device array resonator." Appl. Phys. Lett. **91**, 083509 (2007).

[3] A. R. Matanin, K. I. Gerasimov, E. S. Moiseev, N. S. Smirnov, A. I. Ivanov, E. I. Malevannaya, V. I. Polozov, E. V. Zikiy, A. A. Samoilov, I. A. Rodionov, and S. A. Moiseev, "Toward Highly Efficient Multimode Superconducting Quantum Memory." Phys. Rev. Appl., 19(3), 034011 (2023).

[4] P. K. Day, H. G. LeDuc, B. A. Mazin, A. Vayonakis, and J. Zmuidzinas, "A broadband superconducting detector suitable for use in large arrays." Nature **425**, 6960 (2003).

[5] C. R. H. McRae, H. Wang, J. Gao, M. R. Vissers, T. Brecht, A. Dunsworth, D. P. Pappas and J. Mutus, "Materials loss measurements using superconducting microwave resonators." Rev. Sci. Instrum. **91**, 091101 (2020).

[6] C. Müller, J. H. Cole, and J. Lisenfeld, "Towards understanding two-level-systems in amorphous solids: insights from quantum circuits." Rep. Prog. Phys. **82**, 124501 (2019).

[7] J. M. Martinis, K. B. Cooper, R. McDermott, M. Steffen, M. Ansmann, K. D. Osborn, K. Cicak, Seongshik Oh, D. P. Pappas, R. W. Simmonds and C. Y. Clare, "Decoherence in Josephson qubits from dielectric loss." Phys. Rev. Lett. **95**, 210503 (2005).

[8] W. Woods, G. Calusine, A. Melville, A. Sevi, E. Golden, D. K. Kim, and W. D. Oliver, "Determining interface dielectric losses in superconducting coplanar-waveguide resonators." Phys. Rev. Applied **12**, 014012 (2019).

[9] A. Megrant, C. Neill, R. Barends, B. Chiaro, Y. Chen, L. Feigl, and A. N. Cleland, "Planar superconducting resonators with internal quality factors above one million." Appl. Phys. Lett. **100**, 113510 (2012).

[10] A. Bruno, G. De Lange, S. Asaad, K. L. Van Der Enden, N. K. Langford, and L. DiCarlo, "Reducing intrinsic loss in superconducting resonators by surface treatment and deep etching of silicon substrates." Appl. Phys. Lett. **106**, 182601 (2015).

[11] F. Arute, K. Arya, R. Babbush, D. Bacon, J. C. Bardin, R. Barends, and J. M. Martinis, "Quantum supremacy using a programmable superconducting processor." Nature **574**, 7779 (2019).

[12] S. Kosen, H. X. Li, M. Rommel, D. Shiri, C. Warren, L. Grönberg, and J. Bylander, "Building blocks of a flip-chip integrated superconducting quantum processor." Quantum Sci. Technol. **7**, 035018 (2022).

[13] A. Dunsworth, A. Megrant, C. Quintana, Z. Chen, R. Barends, B. Burkett, and J. M. Martinis, "Characterization and reduction of capacitive loss induced by sub-micron Josephson junction fabrication in superconducting qubits." Appl. Phys. Lett. **111**, 022601 (2017).

[14] C. T. Earnest, J. H. Béjanin, T. G. McConkey, E. A. Peters, A. Korinek, H. Yuan, and M. Mariantoni, "Substrate surface engineering for high-quality silicon/aluminum superconducting resonators." Supercond. Sci. Technol. **31**, 125013 (2018).

[15] C. E. Murray, Mater. Sci. Eng. R Rep. **146**, 100646 (2021).

[16] I. Besedin, and A. P. Menushenkov, "Quality factor of a transmission line coupled coplanar waveguide resonator." EPJ Quantum Technol. **5**, 1 (2018).

[17] I. N. Moskalenko, I. A. Simakov, N. N. Abramov, A. A. Grigorev, D. O. Moskalev, A. A. Pishchimova, N. S. Smirnov, E. V. Zikiy, I. A. Rodionov, and I. S. Besedin, "High fidelity two-qubit gates on fluxoniums using a tunable coupler." npj Quantum Inf., 8(1), 130. (2022).

[18]Z. Chen, A. Megrant, J. Kelly, R. Barends, J. Bochmann, Y. Chen, and J. M. Martinis, "Fabrication and characterization of aluminum airbridges for superconducting microwave circuits." Appl. Phys. Lett. **104**, 052602 (2014).

[19]J. Gao, J. Zmuidzinas, B. A. Mazin, H. G. LeDuc, and P. K. Day, "Noise properties of superconducting coplanar waveguide microwave resonators." Appl. Phys. Lett. **90**, 102507 (2007).

[20]M. Göppl, A. Fragner, M. Baur, R. Bianchetti, S. Filipp, J. M. Fink, and A. Wallraff, "Coplanar waveguide resonators for circuit quantum electrodynamics." J. Appl. Phys. **104**, 113904 (2008).

[21]I. S. Besedin, M. A. Gorlach, N. N. Abramov, I. Tsitsilin, I. N. Moskalenko, A. A. Dobronosova, D. O. Moskalev, A. R. Matanin, N. S. Smirnov, I. A. Rodionov, A. N. Poddubny, and A. V. Ustinov, "Topological excitations and bound photon pairs in a superconducting quantum metamaterial." Phys. Rev. B **103**, 224520 (2021).

[22]D. A. Baklykov, M. Andronic, O. S. Sorokina, S. S. Avdeev, K. A. Buzaverov, I. A. Ryzhikov, and I. A. Rodionov, Micromachines **12**, 534 (2021).

[23]B. B. Lienhard, J. Braumüller, W. Woods, D. Rosenberg, G. Calusine, S. Weber, A. Vepsäläinen, K. O'Brien, T. P. Orlando, S. Gustavsson, W. D. Oliver, "Microwave Packaging for Superconducting Qubits." IEEE MTT-S International Microwave Symposium (IMS) (2019).

[24]R. Barends, J. Wenner, M. Lenander, Y. Chen, R. C. Bialczak, J. Kelly, E. Lucero, P. O'Malley, M. Mariantoni, D. Sank, H. Wang, T. C. White, Y. Yin, J. Zhao, A. N. Cleland, John M. Martinis, and J. J. A. Baselmans, "Minimizing quasiparticle generation from stray infrared light in superconducting quantum circuits." Appl. Phys. Lett. **99**, 113507 (2011).

[25]S. Probst, F. B. Song, P. A. Bushev, A. V. Ustinov, and M. Weides, "Efficient and robust analysis of complex scattering data under noise in microwave resonators." Rev. Sci. Instrum. **86**, 024706 (2015).

[26]J. Gao, "The physics of superconducting microwave resonators." California Institute of Technology (2008).

[27]C. M. Quintana, A. Megrant, Z. Chen, A. Dunsworth, B. Chiaro, R. Barends, and J. M. Martinis, "Characterization and reduction of microfabrication-induced decoherence in superconducting quantum circuits." Appl. Phys. Lett. **105**, 062601 (2014).

[28]J. Wenner, R. Barends, R. C. Bialczak, Y. Chen, J. Kelly, E. Lucero, and J. M. Martinis, "Surface loss simulations of superconducting coplanar waveguide resonators." Appl. Phys. Lett. **99**, 113513 (2011).

[29]A. Melville, G. Calusine, W. Woods, K. Serniak, E. Golden, B. M. Niedzielski, and W. D. Oliver, "Comparison of dielectric loss in titanium nitride and aluminum superconducting resonators." Appl. Phys. Lett. **117**, 124004 (2020).

[30]A. Nersisyan, S. Poletto, N. Alidoust, R. Manenti, R. Renzas, C. V. Bui, and M. Reagor, "Manufacturing low dissipation superconducting quantum processors." IEEE International Electron Devices Meeting (IEDM) (2019).

[31]J. M. Sagea, V. Bolkhovsky, W. D. Oliver, B. Turek, and P. B. Welander, "Study of loss in superconducting coplanar waveguide resonators." J. Appl. Phys. **109**, 063915 (2011).

[32]I. A. Rodionov, A. S. Baburin, A. R. Gabidullin, S. S. Maklakov, S. Peters, I. A. Ryzhikov, and A. V. Andriyash, "Quantum engineering of atomically smooth single-crystalline silver films." Sci. Rep. **9**, 12232 (2019).

[33]G. P. Fedorov, S. V. Remizov, D. S. Shapiro, W. V. Pogosov, E. Egorova, I. Tsitsilin, M. Andronik, A. A. Dobronosova, I. A. Rodionov, O. V. Astafiev, and A. V. Ustinov, "Photon transport in a Bose-Hubbard chain of superconducting artificial atoms." Phys. Rev. Lett. **126**, 180503 (2021).

# Supplement to: High-Q trenched aluminum coplanar resonators with an ultrasonic edge microcutting for superconducting quantum devices

E.V. Zikiy,[1,2] A.I. Ivanov,[1,2] N.S. Smirnov,[1,2] D.O. Moskalev,[1,2]
V.I. Polozov,[1] A.R. Matanin,[1,2] E.I. Malevannaya,[1] V.V. Echeistov,[1]
T.G. Konstantinova[1] and I.A. Rodionov[1,2]

[1]FMN Laboratory, Bauman Moscow State Technical University, Moscow 105005, Russia

[2]Dukhov Automatics Research Institute (VNIIA), Moscow 127055, Russia

This supplement provides experimental details and data sets to support the claims made in the main text. First, we present the design and fabrication details for the two types of devices we investigated: a resonator circuit without substrate etching and a resonator circuit with deep substrate etching. We then describe a several technical details of the measurement system: the measurement setup and device shielding and the fitting of the complex-valued transmission spectra. Next, we present the measurement results for the type 1 and type 2 devices.

## DEVICES

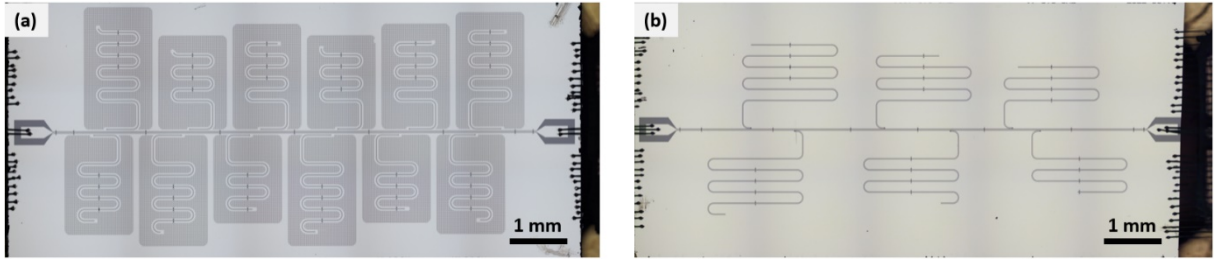

FIG. S1: (a) Optical image of a 12-resonators circuit with frequencies from 4.0 to 7.0 GHz without substrate etching (type 1). (b) Optical image of a 6-resonators circuit with frequencies from 2.5 to 5.0 GHz with isotropic substrate etching (type 2). Both resonator circuits are fabricated using wet etching in an industrial Aluminum Etching Type A solution. Resonator circuits with airbridges over feedline and resonators are presented.

On the chip without substrate etching (Fig. S1(a)) there are 12 resonators with frequencies: 4.0, 4.4, 4.6, 4.8, 5.0, 5.4, 5.6, 6.0, 6.4, 6.6, 6.8, 7.0 GHz and on the chip with isotropic substrate etching (Fig. S1(b)) there are 6 resonators with frequencies: 2.5, 3.0, 3.5, 4.0, 4.5, 5.0 GHz. For silicon isotropic etching we use hight-density plasma etching at RF power of 700 W and bias power of 40 W, a $SF_6$ with 35 sccm flow rates, and 40 mTorr pressure. Table S1 shows the fabrication parameters of all devices presented in this work.

## MEASUREMENT SETUP

To minimize resonator losses induced by non-equilibrium quasiparticles and magnetic vortex displacement, we mount the samples inside several layers of shielding (Fig. S2). Specifically, we anchor the PCB-mounted sample directly to a copper cold finger connected to the mixing chamber of the dilution refrigerator. The sample is then enclosed in a copper can, the inner surface of which is coated in a mixture of Stycast 2850 FT and silicon carbide granules with diameter 1000 μm. This can is enclosed in a second aluminum can. This is finally enclosed in one layer of cryogenic magnetic shielding (1-mm-thick Cryoperm). Coaxial cables entering the sample holder were pasted into the lid of the inner layer of IR shielding to reduce the impact of open holes on the shielding effectiveness. Extra radiation shielding is provided by in-house inline Eccosorb infrared filters in the input coaxial line mounted outside the magnetic shields at the mixing chamber stage.

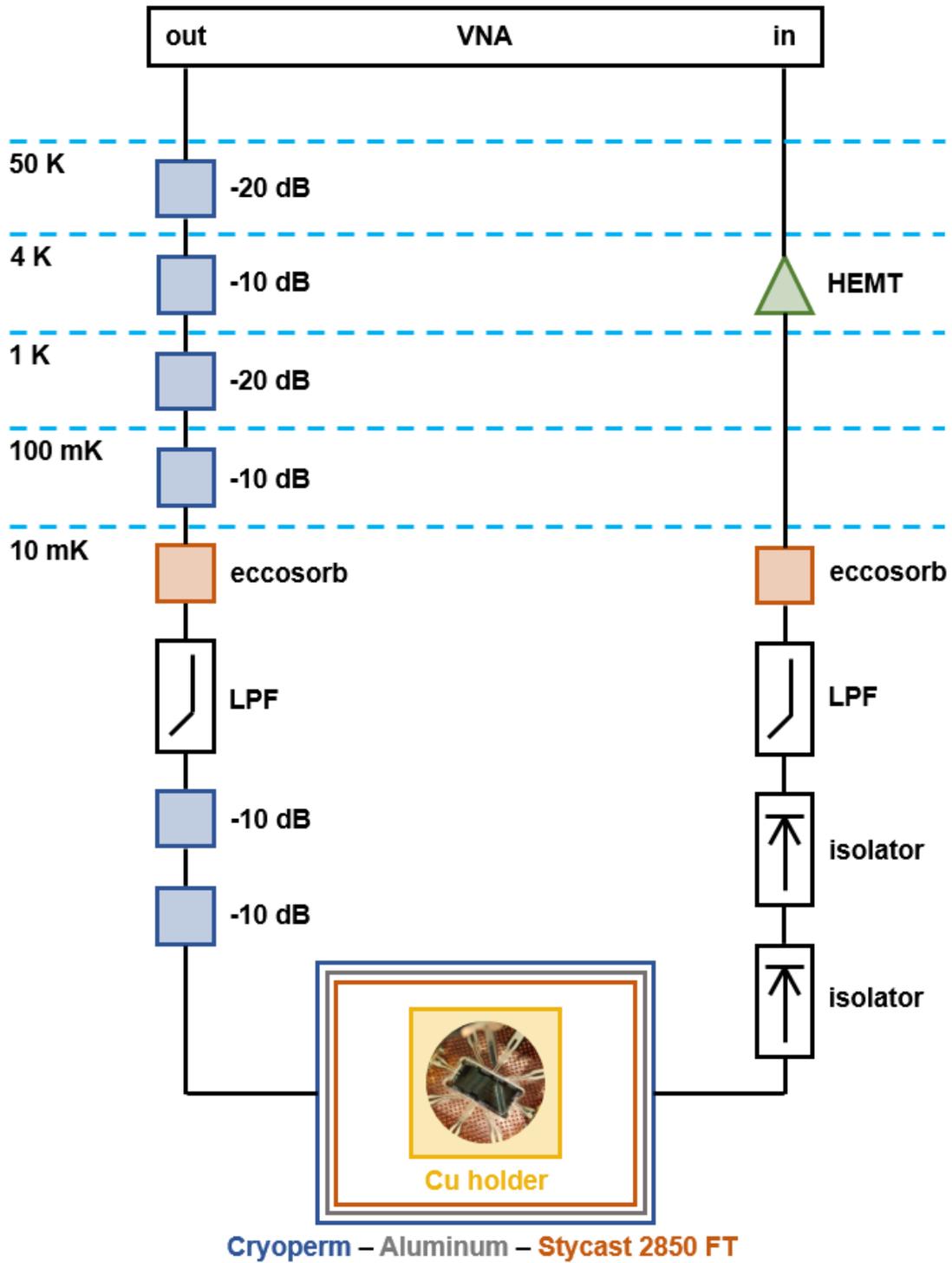

FIG. S2: Wiring diagram of a measurement setup for the samples

TABLE SI. Device fabrication parameters

| device | f, GHz | $Q_{iLP} \times 10^6$ | $Q_{iHP} \times 10^6$ | wet etch Al | RIE Al | Bosch Si | iso. etch Si | US microcutting | feedline bridges | resonator bridges |
|---|---|---|---|---|---|---|---|---|---|---|
| 1a.1 | 4.45 | 0.718 | 2.739 | | x | | | | | |
| 1a.2 | 4.55 | 0.619 | 2.614 | | x | | | | | |
| 1a.3 | 4.84 | 0.674 | 5.292 | | x | | | | | |
| 1a.4 | 4.45 | 0.435 | 2.206 | | x | | | | | |
| 1a.5 | 4.55 | 0.591 | 3.588 | | x | | | | | |
| 1a.6 | 4.94 | 0.490 | 3.089 | | x | | | | | |
| 1a.7 | 4.44 | 0.638 | 5.852 | | x | | | | | |
| 1a.8 | 4.53 | 0.634 | 5.614 | | x | | | | | |
| 1b.1 | 4.40 | 1.193 | 4.864 | x | | | | | | |
| 1b.2 | 4.40 | 1.226 | 4.797 | x | | | | | | |
| 1b.3 | 4.60 | 0.947 | 0.947 | x | | | | | | |
| 1b.4 | 4.60 | 1.215 | 3.059 | x | | | | | | |
| 1b.5 | 4.40 | 1.406 | 5.095 | x | | | | | | |
| 1b.6 | 4.60 | 1.128 | 4.395 | x | | | | | | |
| 1b.7 | 4.82 | 1.182 | 3.646 | x | | | | | | |
| 1b.8 | 4.46 | 1.144 | 6.032 | x | | | | | | |
| 2a.1 | 4.56 | 0.667 | 2.623 | x | | x | | | | |
| 2a.2 | 4.55 | 0.593 | 4.122 | x | | x | | | | |
| 2b.1 | 4.39 | 0.862 | 2.648 | x | | | x | | | |
| 2b.2 | 4.02 | 0.776 | 2.791 | x | | | x | | | |
| 2b.3 | 4.69 | 1.417 | 2.945 | x | | | x | | | |
| 2a.3 | 4.55 | 0.525 | 3.569 | x | | x | | x | | |
| 2a.4 | 4.04 | 0.673 | 3.887 | x | | x | | x | | |
| 2a.5 | 4.64 | 0.641 | 3.861 | x | | x | | x | | |
| 2a.6 | 4.04 | 0.557 | 5.367 | x | | x | | x | | |
| 2b.4 | 4.07 | 1.933 | 6.117 | x | | | x | | | |
| 2b.5 | 4.32 | 1.097 | 2.251 | x | | | x | | | |
| 2b.6 | 4.07 | 1.172 | 3.816 | x | | | x | | | |
| 2c.1 | 4.33 | 2.005 | 5.820 | x | | | x | x | | |
| 2c.2 | 4.36 | 2.346 | 7.100 | x | | | x | x | | |
| 2c.3 | 4.72 | 1.645 | 5.169 | x | | | x | x | | |
| 2c.4 | 4.00 | 2.184 | 10.391 | x | | | x | x | | |
| 2c.5 | 2.91 | 4.447 | 18.297 | x | | | x | x | | |
| 2c.6 | 3.25 | 3.457 | 13.407 | x | | | x | x | | |
| 2c.7 | 4.39 | 1.861 | 7.203 | x | | | x | x | x | |
| 2c.8 | 4.02 | 1.680 | 9.176 | x | | | x | x | x | |
| 2c.9 | 4.76 | 1.884 | 6.159 | x | | | x | x | x | |
| 2c.10 | 4.02 | 1.388 | 5.371 | x | | | x | x | x | x |
| 2c.11 | 4.39 | 1.095 | 2.335 | x | | | x | x | x | x |
| 2c.12 | 4.76 | 1.238 | 2.957 | x | | | x | x | x | x |